\documentclass[prb,groupedaddress,showpacs,twocolumn]{revtex4}
\usepackage[dvips]{graphicx}

\begin{document}
\title{The possible explanation of electric-field-doped C$_{60}$ phenomenology in the framework of Eliashberg theory}
\author{G.A. Ummarino, R.S. Gonnelli}
\affiliation{$INFM-$Dipartimento di Fisica, Politecnico di Torino,
Corso Duca degli Abruzzi, 24-10129 Torino, Italy\\ E-mail address:
ummarino@polito.it}
\date{\today}
\begin{abstract}
%\vspace{5mm}
In a recent paper (J.H. Sch$\ddot{o}$n, Ch. Kloc, R.C. Haddon and
B. Batlogg, Nature 408 (2000) 549) a large increase in the
superconducting critical temperature was observed in C$_{60}$
doped with holes by application of a high electric field. We
demonstrate that the measured $T_\mathrm{c}$ versus doping curves
can be explained by solving the (four) s-wave Eliashberg equations
in the case of a finite, non-half-filled energy band. In order to
reproduce the experimental data, we assume a Coulomb
pseudopotential depending on the filling in a very simple and
plausible way. Reasonable values of the physical parameters
involved are obtained. The application of the same approach to new
experimental data (J.H. Sch$\ddot{o}$n, Ch. Kloc and B. Batlogg,
Science 293 (2001) 2432) on electric field-doped, lattice-expanded
C$_{60}$ single crystals ($T_\mathrm{c}=$117 K in the hole-doped
case) gives equally good results and sets a theoretical limit to
the linear increase of $T_\mathrm{c}$ at the increase of the
lattice spacing.
 %\vspace{5mm}
\end{abstract}
 \maketitle
%\narrowtext
After obtaining the surface field-effect doping of high-quality
organic crystals, J.H. Sch$\ddot{o}$n, Ch. Kloc, R.C. Haddon and
B. Batlogg have carried out the same experiment on C$_{60}$
crystals \cite{refe1,refe2,refe3}, achieving superconductivity up
to 52 K in hole-doped samples \cite{refe1}. The fundamental
importance of this experiment is due to the fact that the
superconductivity is obtained by field-effect doping the original
single-crystal material and, thus, without modifying its
structural properties. The authors of the experiments believe as
plausible that the doping charges are confined only in a single
layer of C$_{60}$ molecules \cite{refe1}.
 The physical parameters of C$_{60}$ are rather well known; the symmetry of the order parameter
 appears to be \emph{s}-wave and the electron-phonon interaction is widely indicated as the main cause for
the appearance of superconductivity in this material. In the
present paper we show that a theory suitable for reproducing the
experimental $T_\mathrm{c}$ versus filling data reported in Ref.
1 is the Migdal-Eliashberg theory
\cite{refe4,refe5,refe6,refe6a,refe7,refe8,refe9}, even though, in
fullerenes, the Migdal Theorem probably breaks down \cite{refe10}
due to the large energy of phonons and the rather low energy of
the Fermi level \cite{refe11,refe11a}.
 As a first approximation, we neglect these complications and see
 where the theory in its most simple form and with the smallest number
of free parameters can lead. Notice that a study of the
phenomenology of field-doped C$_{60}$ in the framework of the
Eliashberg theory was carried out also in Ref. 14 although in a
different way.
\begin{figure}[t]
 \vspace{-1mm}
 \includegraphics[keepaspectratio,width=0.9\columnwidth]{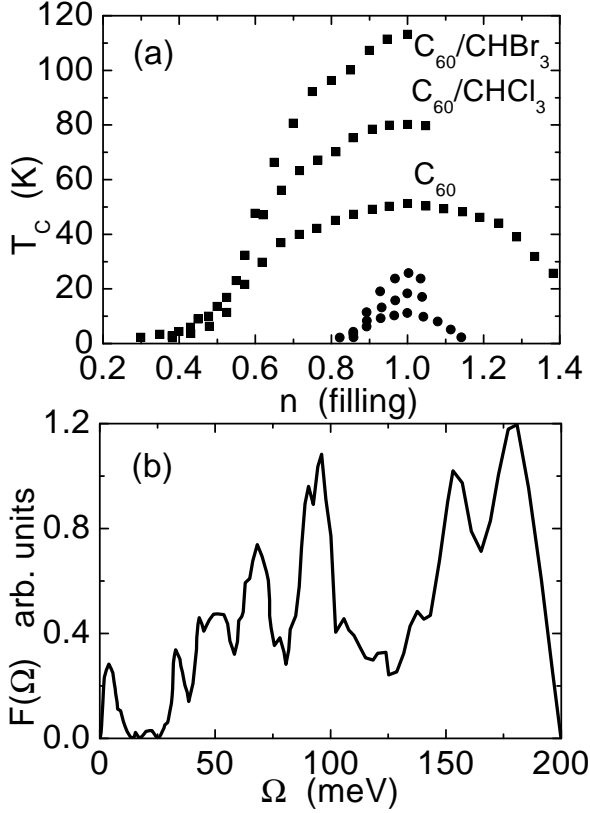}
 \vspace{-2mm}\caption{\small{(a) Experimental $T_{c}$ versus
 filling $n$ data for hole doping (full squares) and electron doping (full circles) [from \cite{refe1,refe13}];
 (b) The C$_{60}$ phonon density of states $F(\Omega)$ [from \cite{refe11}].}}\vspace{-5mm}
 \end{figure}
 Figure 1 (a) was obtained by normalizing the original data from
Ref. 1 (and from more recent experimental data in lattice-expanded
C$_{60}$, as we will discuss later) so that the maximum of
$T_{\mathrm{c}}$ corresponds to half filling (here indicated by
$n=1$). In order to fit these experimental data, we solved the
Eliashberg equations in the general case of non-half filling,
finite bandwidth and flat normal density of states. The physical
quantities that appear in the fit are thus: the bandwidth W, the
electron-phonon spectral function $\alpha^{2}F(\Omega)$ and the
Coulomb pseudopotential $\mu$. It is also necessary to set a
cut-off energy $\omega_{C}$ related to the renormalization of the
Coulomb pseudopotential. We took the value of W from literature
\cite{refe11} : W=250 meV (independent of doping since the
lattice structure does not change \cite{refe1}) and we assumed
the cut-off energy $\omega_{C}$=W as in Ref. 10. The spectral
function is unknown but, as a first approximation, one can assume
that $\alpha^{2}F(\Omega)$ is proportional to the phonon density
of states \cite{refe11} $F(\Omega)$ of C$_{60}$ (see figure 1
(b)) and nearly equal to the transport spectral function
$\alpha^{2}F_{tr}(\Omega)$.

To determine the proportionality factor between
$\alpha^{2}F(\Omega)$ and $F(\Omega)$, we took the experimental
resistivity data $\rho(T)$ for the electron-doping case from the
figure 5 of Ref. 1, in the range between $T = 225$ K and $T= 245$
K. The $\rho(T)$ curves appear to have the same shape and slope
at all gate voltages, thus, as a first approximation, the ratio
$\lambda_{tr}/\omega_{P}$ between the transport coupling constant
$\lambda_{tr}$ and the plasma frequency $\omega_{P}$ can be
assumed to be doping-independent. The phonon density of states
$F(\Omega)$ and the functions $\alpha^{2}_{e,h}(\Omega)$ (for
both electrons and holes) are doping-independent as well, since
the lattice parameters do not change \cite{refe6}. As a result,
also the electron-phonon coupling constants $\lambda_{e,h} = 2
\int \mathrm{d}\Omega \alpha^2_{e,h} F(\Omega)/\Omega$ are
independent of doping. From the high-temperature expression of
the resistivity: $\rho= 8 \pi k_{B} T
\lambda_{tr}/(\omega^{2}_{P} \varepsilon_{0}\hbar^{2})$, and by
using the value of the plasma frequency reported in literature
\cite{refe11} $\hbar\omega_{P}=1.36$ eV, we can deduce the value
of the transport electron-phonon coupling constant $\lambda_{tr}$
\cite{refe11ab}. In the electron-doping case we find
$\lambda_{tr}= 1.1$ and, as a first approximation, we assume
$\lambda_{tr}\equiv \lambda_{e}$. As far as the hole-doped
coupling constant $\lambda_{h}$ is concerned, from resistivity
data and from calculations of the electron density of states J.H.
Sch\"{o}n et al. \cite{refe2} estimate that $\lambda_{h}/
\lambda_{e}=1.5$ and so we find $\lambda_{h}=1.5\lambda_{e}=1.65$.

 In order to obtain the exact maximum value of the critical temperature
in the half-filling case, as shown in figure 1 (a), we solved the
Eliashberg equations in the half-filling case by using for the
Coulomb pseudopotential $\mu$ the values 0.395 and 0.336 in the
electron and hole doped case, respectively. This result is in very
good agreement with theoretical calculations \cite{refe11} that
predict $\mu=0.3-0.4$.

Two critical remarks can be made to our basic assumptions.

 First: due to the degeneracy of $C_{60}$-orbitals, one has probably
to deal with a multi-band problem \cite{refe11b}. It is well
known that, in such a case, transport and superconducting
properties can be managed by different groups of electrons but,
in this way, there would be several free parameters and the model
would become too complex. Instead, our simple approximation allows
precise quantitative predictions.

Second: we have assumed for simplicity that the normal density of
states (NDOS) is flat even if in recent theoretical calculations
\cite{refe11c} it is not so. This calculated NDOS can be
approximately reproduced by a very simple analytical formula:
$N_{N}(\omega)=1+1.5exp(-|\omega|/\alpha)$ with $\alpha= $ 10
meV. If we use this expression of the NDOS in the Eliashberg
equations, with the same values of $\mu$, i.e. 0.395 and 0.336,
we find that the exact maximum $T_\mathrm{c}$ is obtained by
using $\lambda=0.843$  and $\lambda=1.58$. The difference with
respect to the case where $N_{N}(\omega)=1$ is small, so that the
I-V curves measured in the tunable SNS weak-link junctions of Ref.
3 can be fitted equally well by both the NDOS.
 Moreover, if the energy dependence of the NDOS is symmetric and the peak at the Fermi level
 is not too narrow, its effects on some physical quantities can be
approximately simulated by an efficient value \cite{refe6} of
$\lambda$. In conclusion, if $\Delta E=(\omega^{2}_{P}
\varepsilon_{0}\hbar^{2})\rho_{0}/7.5 \geq \hbar\omega_{D}$ (where
$\omega_{D}$ is the Deybe frequency and $\rho_{0}$ is the residual
resistivity), the effect of a non-flat NDOS can be neglected
\cite{refe6}. In our case \cite{refe1} $\rho_{0}=250\div300$
$\mu\Omega$ cm, $\Delta E=62\div74 $ meV  and
$\hbar\omega_{D}\cong \hbar\omega_{ln}=27 $ meV, where $\omega
_{\log }=\exp(\frac{2}{\lambda }\int_{0}^{+\infty }d\Omega
\frac{\alpha ^{2}F(\Omega)}{\Omega }\ln \Omega)$.

 Since the lattice is not modified by the field-effect doping, we
assume the electron-phonon coupling constant $\lambda$ and the
width of the conduction band W independent of filling $n$.
  Now we can try to reproduce the experimental values of $T_\mathrm{c}$
versus $n$ by solving the Eliashberg equations in the
non-half-filling case \cite{refe7}. This is a hard task because
there are three equations to be solved for the calculation of the
gap $\Delta(i\omega_{n})$, the renormalization function
$Z(i\omega_{n})$ and the asymmetric part of the self-energy
$\chi(i\omega_{n})$ -- which is always equal to zero in the
half-filling case -- plus one equation that represents the
conservation of the particles' number and is necessary for
calculating the shift of the chemical potential $\delta\mu$. Thus
the Eliashberg equations in the non-half-filling case, for finite
bandwidth and \emph{s}-wave symmetry of the order parameter read:
\cite{refe7}:
%\vspace{10mm}
\begin{eqnarray}
\Delta(i \omega_{n})Z(i \omega_{n})&=&\pi k_{B}T
\sum_{m=-\infty}^{+\infty}\left[\lambda(i \omega_{n}-i
\omega_{m})-\mu^{*}/W\right]\cdot \nonumber
\\ & &\vartheta(|\omega_{C}|-\omega_{m})P(i
\omega_{m})\Theta(i \omega_{m})
\end{eqnarray}
%\vspace{10mm}
\begin{equation}
\omega_{n}Z(i\omega_{n})=\pi k_{B}T\sum_{m=-\infty}^{+\infty}
\lambda(i\omega_{n}-i\omega_{m})N(i\omega_{m})\Theta(i\omega_{m})
\end{equation}
%\vspace{10mm}
\begin{equation}
\chi(i\omega_{n})=-\pi k_{B}T\sum_{m=-\infty}^{+\infty}
\lambda(i\omega_{n}-i\omega_{m})M(i\omega_{m})
\end{equation}
%\vspace{10mm}
\begin{equation}
 n=1-(\pi k_{B}T/W)\sum_{m=-\infty}^{+\infty}M(i\omega_{m})
\end{equation}
 %\vspace{10mm}
where $\vartheta$ is the Heaviside function,
\makebox{$i\omega_{m}=i\pi(2m+1)k_{B}T$} with $m=0,\pm1,\pm2,...$
and
 %\vspace{10mm}
\begin{equation}
\vspace{-3mm}
 \mu^{*}=\mu/[1+\frac{k_{B}T \mu}{W}\int^{W}_{-W}d\varepsilon \Sigma_{m=0}^{+\infty}
\frac{\vartheta(\omega_{m}-\omega_{C})}{\omega_{m}^{2}+(\varepsilon-\delta\mu)^{2}}]
\end{equation}
%\vspace{10mm}
\begin{equation}
\vspace{-3mm}
\lambda(i\omega_{n}-i\omega_{m})=\int_{0}^{+\infty}d\Omega\frac{\alpha^{2}(\Omega)F(\Omega)}{\Omega^{2}+(i\omega_{n}-i\omega_{m})^{2}}
\end{equation}
%\vspace{18mm}
\begin{equation}
P(i\omega_{m})=\frac{\Delta(i\omega_{m})Z(i\omega_{m})}{R(i\omega_{m})},
    N(i\omega_{m})=\frac{\omega_{m}Z(i\omega_{m})}{R(i\omega_{m})}
\end{equation}
\begin{equation}
R(i\omega_{m})=\sqrt{\omega^{2}_{m}Z^{2}(i\omega_{m})+\Delta^{2}(i\omega_{m})Z^{2}(i\omega_{m})}
\end{equation}
%\vspace{30mm}
\begin{equation}
\Theta(i\omega_{m})=\{\arctan(W_{-}(i\omega_{m}))+\arctan(W_{+}(i\omega_{m}))\}/\pi
\end{equation}
\begin{equation}
M(i\omega_{m})=0.5\ln[\frac{1+W^{2}_{-}(i\omega_{m})}{1+W^{2}_{+}(i\omega_{m})}]
\end{equation}
and
$W_{\mp}(i\omega_{m})=\{W\mp[\delta\mu-\chi(i\omega_{m})]\}/R(i\omega_{m})$.

 \begin{figure}[!]
 \vspace{-1mm}
 \includegraphics[keepaspectratio,width=0.9\columnwidth]{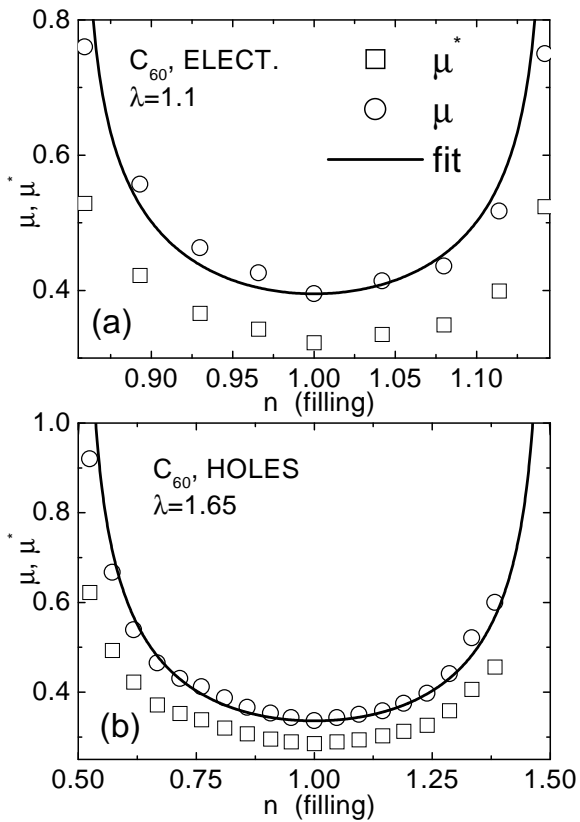}
 \vspace{-2mm}\caption{\small{Coulomb pseudopotential $\mu$ (open
  circle), $\mu^{*}$ (open square) and theoretical fit (solid line) versus filling n: (a)
  electron-doped case, (b) hole-doped case.}}\vspace{-5mm}
 \end{figure}
 In order to reproduce the experimental values of $T_\mathrm{c}$ versus doping, and having
 excluded the dependence of $\lambda$ on the filling,
  the only possibility is to assume that the Coulomb
pseudopotential $\mu$ is variable with the filling $n$:
$\mu\equiv\mu(n)$. We find the values of $\mu(n)$ by solving the
equations (1)-(4) conditioned to the obtainment of the
experimental $T_\mathrm{c}(n)$. In figure 2 we show the $\mu(n)$
dependence in the two cases: (a) electron-doped C$_{60}$ and (b)
hole-doped C$_{60}$ (open circles). It is important to notice that
here the definition of $\mu^{*}$ is more general than usual
because it depends on the shift of the chemical potential and on
the value of the bandwidth. In the usual and less general
definition \cite{refe5}, if $\omega_{C}=W$ then $\mu=\mu^{*}$
while in our case it is always $\mu>\mu^{*}$ as we can see in
figure 2 (open squares).

  Is it possible to explain this particular dependence of $\mu$ on
  the filling, obtained by the fit of the critical temperature of figure
1 (a)? This can be done in two or three dimensions. In a very
recent paper \cite{refe12aa} the authors affirm that it is not so
sure that the superconductivity in this material is 2D. For
completeness we examine both the 2D and 3D cases. From the
definition of $\mu$ \cite{refe12} and from the very simple
analytical expression of the Thomas-Fermi dielectric function in
the 3D case
 $\varepsilon(q,n)=1+k^{2}_{S}(n)/q^{2}$ it follows:
  %\vspace{5mm}
  \begin{equation}
\mu(n)=\frac {1}{4\pi^{2}\hbar v_{F}}
\int_{0}^{2k_{F}}\frac{V(q)}{\varepsilon(q,n)} qdq
\end{equation}
%\vspace{5mm}
where $V(q)=4\pi e^{2}/q^{2}$.
 In the half-filling case, the calculation of the integral gives:
 %\vspace{5mm}
   \begin{equation}
   \mu(n=1)=[\frac{k^{2}_{S}(n=1)}{8
   k^{2}_{F}}]\ln[1+(\frac{4k^{2}_{F}}{k^{2}_{S}(n=1)})]
   \end{equation}
    %\vspace{5mm}
 where $k_{F}$ and $k_{S}$ are the Fermi and Thomas-Fermi wave vector, respectively.
 We can expand the dielectric function in the
vicinity of the half filling ($n=1$) by remembering that, since
the physical parameters $\lambda $, $\alpha^{2}F(\Omega)$ and W
are fixed, $T_\mathrm{c}$ is maximum when $\mu(n)$ is minimum,
which means that $ \varepsilon(n)$ is maximum (see definition of
$\mu(n)$) i.e. $\partial \varepsilon /\partial n =0 $ for $n=1$.
For simplicity, and to minimize the number of free parameters, we
arrest the expansion to the second-order terms, even though the
range of the series expansion is not so small in the hole-doping
case:
\begin{equation}
\varepsilon(q,n)=\varepsilon(q,n=1)+\frac{1}{2}[\frac{\partial^{2}\varepsilon(q,n)}{\partial
n^{2}}]_{n=1}(n-1)^{2}+...
\end{equation}
and by substituting in the definition of $\mu(n)$ we find
\begin{equation}
\mu(n)=\mu(n=1)\frac{ln(1+a^{2}/[1+b(n-1)^{2}])}{ln(1+a^{2})}
\end{equation} where
 $a=2k_{F}/k_{S}(n=1)$ and $b=[(1/
k_{S})(\partial^{2} k_{S}/ \partial n^{2})]_{n=1}$. By starting
from the values of $\mu(n=1)$ that give the exact experimental
$T_\mathrm{c}$ in the electron- and hole-doping case and using eq.
12 we can calculate $2 k_{F}/ k_{S}(n=1)$. We obtain 1.05 for hole
doping and 0.77 for electron doping. Now we are left with only one
free parameter, $b=(1/ k_{S})[\partial^{2} k_{S}/
\partial n^{2}]_{n=1}$, that can be adjusted to fit
the $\mu$ versus $n$ curves with eq. 14. The results are very
good (see figure 2 (a) and (b), solid lines): for electron doping
$b=-25$ and for hole doping $b=-4$. Of course, here we used the
3D version of the Thomas-Fermi theory but, as pointed out by the
authors of ref. 1-3, the field-effect charge injections is likely
to be confined to the first atomic layer of $C_{60}$ crystal, so
it is worthwhile to examine also the 2D case. The analytical
expression of the Thomas-Fermi dielectric function \cite{refe12a}
is now $\varepsilon(q,n)=1+k_{S}(n)/q$ and
  %\vspace{5mm}
  \begin{equation}
\mu(n)=\frac {k_{F}}{\pi\hbar v_{F}}
\int_{0}^{2k_{F}}\frac{V(q)}{\varepsilon(q,n)\sqrt{4k^{2}_{F}-q^{2}}}dq
\end{equation}
%\vspace{5mm}
where $V(q)=2\pi e^{2}/q$ \cite{refe12a}. By following the same
approach as before we find:
\begin{equation}
\mu(n=1)=2\frac{[\arctan(\frac{a}{\sqrt{a^{2}-1}})-1]}{\sqrt{a^{2}-1}}
\end{equation}
and finally
%
%\begin{equation}
%\mu(n)=\mu(n=1)\frac{\sqrt{a^{2}-1}}{\sqrt{a^{2}[1+0.5b(n-1)^{2}]^{2}-1}}}
%\cdot\frac{[\arctan(\frac{a[1+0.5b(n-1)^{2}]}{\sqrt{a^{2}[1+0.5b(n-1)^{2}]^{2}}})-1]}{
%[\arctan(\frac{a}{\sqrt{a^{2}-1}})-1]}
%\end{equation}
\begin{equation}
\mu(n)=\mu(n=1)\frac{\sqrt{a^{2}-1}}{\sqrt{a^{*2}-1}}
\cdot\frac{[\arctan(\frac{a^{*}}{\sqrt{a^{*2}-1}})-1]}{
[\arctan(\frac{a}{\sqrt{a^{2}-1}})-1]}
\end{equation}
with $a^{*}=a[1+0.5b(n-1)^{2}]$.

In this case, from eq. 16, we can calculate $a\equiv2 k_{F}/
k_{S}(n=1)$. We find two solutions: 1.9 and 13.8 for hole doping
and 2.1 and 10.3 for electron doping. Again we have only one free
parameter, $b=(1/ k_{S})[\partial^{2} k_{S}/
\partial n^{2}]_{n=1}$, that can be adjusted to fit the $\mu$ versus $n$ curves with eq. 17.
We find $b=-43$ and $b=-73$ for electron doping while for hole
doping we have $b=-4$ and $b=-8$, since now two possible values
for $a$ are present. In the range of interest, the results are
exactly equal to the 3D case and the 2D curves perfectly overlap
those obtained in that case. If we compare the 2D and 3D
densities of charge carriers obtained from the $a$ values of the
fits to the values predicted at optimal doping and corresponding
to 3 carriers per C$_{60}$ ($\sim 3\cdot10^{18}$ carriers/m$^2$
and $\sim 4.2\cdot10^{27}$ carriers/m$^3$, respectively) we obtain
interesting results. The 3D density and the 2D one obtained from
the small $a$ values are of the same order of magnitude of the
predicted densities, even if they don't coincide. On the contrary
the 2D density obtained from the large $a$ values is more than 2
orders of magnitude greater than the predicted one and, thus, the
corresponding solution for $a$ must be rejected on the basis of
physical arguments. It follows that the results of the fit by
using our model do not allow us to understand whether the system
is better described by 2D or 3D approach, even if the
experimental results \cite{refe1, refe13} clearly point to a 2D
nature of the charge injected region.

 Note that, surprisingly, the curves T$_{c}(n)/T_{c}(n=1)$ and
$\mu(n=1)/\mu(n)$ are perfectly superimposed.

Now, since we have the approximate analytical dependence of $\mu$
from $n$, we can solve the Eliashberg equations for any value of
the filling and calculate different physical quantities.
\begin{figure}[t]
 \vspace{-1mm}
 \includegraphics[keepaspectratio,width=0.9\columnwidth]{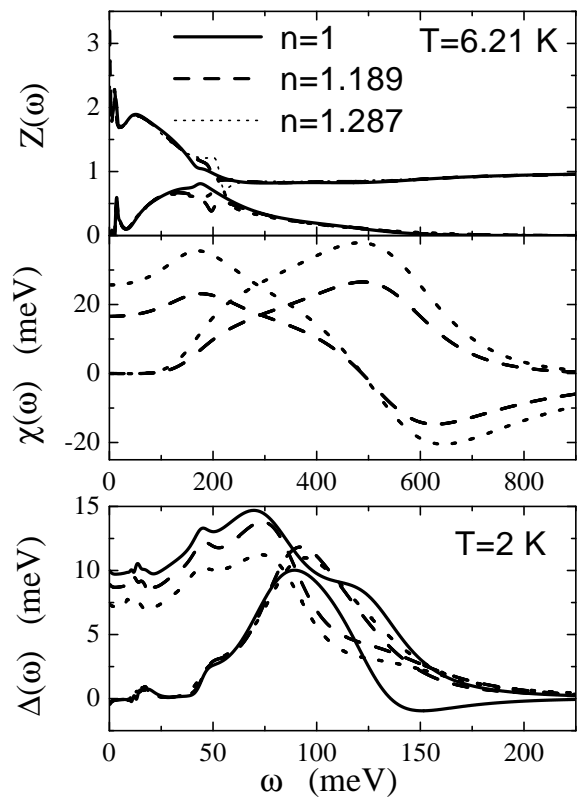}
 \vspace{-1mm}\caption{\small{The real and imaginary part of the functions $Z(\omega)$,
 $\xi(\omega)$ at $T=6.21$ K and of $\Delta(\omega)$ at $T=2$ K for different
values of the filling:
  $n=1$ (solid line), $n=1.189$ (dash), and $n=1.287$ (dot) in the hole-doped case.}}\vspace{-5mm}
 \end{figure}
 \begin{figure}[!]
 \vspace{5mm}
 \includegraphics[keepaspectratio,width=0.9\columnwidth]{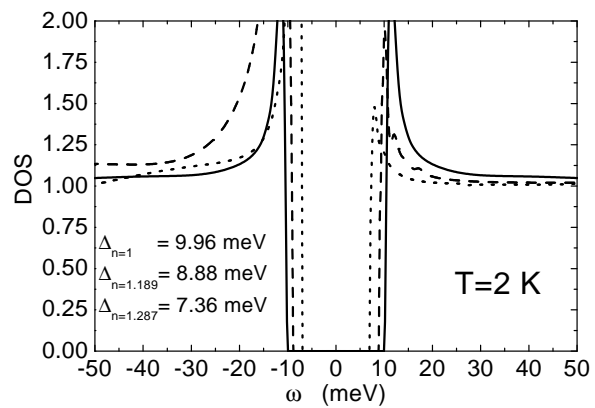}
 \vspace{-1mm}\caption{\small{The normalized DOS, at T=2 K, for different
values of the filling: $n=1$ (solid line), $n=1.189$ (dash), and
$n=1.287$ (dot) in the hole-doped case.}}\vspace{-1mm}
 \end{figure}
By using the standard technique of Pad\'{e} approximants we can
carry out the analytical continuation of $Z(i\omega_{n})$,
$\Delta(i\omega_{n})$, $\chi(i\omega_{n})$ and of the normalized
superconductive density of states (DOS)
$N_{S}(i\omega_{n})/N_{N}(i\omega_{n})=[N(i\omega_{n})\Theta(i\omega_{n})-M(i\omega_{n})]/[N_{N}(i\omega_{n})\Theta_{N}(i\omega_{n})-M_{N}(i\omega_{n})]$.
Here the quantities with suffix N are calculated for
$\Delta(i\omega_{n})=0$. In figure 3, we can see how the shape of
$Z(\omega)$, $\Delta(\omega)$ and $\chi(\omega)$ changes for three
different values of the doping in the hole-doped case: $n=1$
($T_{c}=51$ K), $n=1.189$ ($T_{c}=46$ K) and $n=1.287$ ($T_{c}=39$
K). Figure 4 shows the normalized superconducting DOS as a
function of the energy, calculated by the analytical continuation
of the imaginary-axis DOS for the same doping values as in Fig.
3. In the non-half-filling cases ($n=1.189$ and $n=1.287$) the
calculated DOS is asymmetric. In principle low-temperature SIN
tunneling experiments on this field-doped material would make it
possible to observe this asymmetry. The calculated values of the
ratio $2\Delta/k_{B}T_{c}$ are slightly greater than the
experimental results of Ref. 3.

Very recently J.H. Sch\"{o}n \emph{et al}. \cite{refe13} have
intercalated single crystals of C$_{60}$ with CHCl$_{3}$ and
CHBr$_{3}$ in order to expand the lattice (see figure 1). The
maximum of the critical temperature (onset of resistive
transition) in the hole-doped C$_{60}$/CHBr$_{3}$ is 117 K. We
can try to explain these new experimental data by following the
approach previously illustrated. We note that the maximum of
T$_{c}$ occurs at the same value of doping in the three
hole-doped and electron-doped cases (see figure 1 (a)).

As a first approximation we can assume that the Coulomb
pseudopotential is, for $n=1$, the same as in the bare C$_{60}$.
For simplicity, we also use the electron-phonon spectral function
of C$_{60}$ simply multiplied by a constant in order to obtain
the correct $T_{c}$. These two approximations can be justified by:
(i) the presence of an almost constant density of charge carriers
\cite{refe1,refe13} in the bare and intercalated C$_{60}$; (ii)
the weak dependence of $T_{c}$ from the details of the actual
form of the spectral function \cite{refe20} ($T_{c}$ strongly
depends on $\omega _{\log }$ and on the total energy range of the
spectral function that are likely not to be changed very much by
the intercalation). The values of the coupling constant $\lambda$
necessary to get the new experimental critical temperatures are
2.11 and 2.76 in the hole-doped case ($T_{c}=80$ K and
$T_{c}=114$ K, respectively)
 and 1.24 and 1.37 in the electron-doped case ($T_{c}=18$ K and $T_{c}=26$ K, respectively).
 \begin{figure}[!]
 \vspace{-1mm}
 \includegraphics[keepaspectratio,width=\columnwidth]{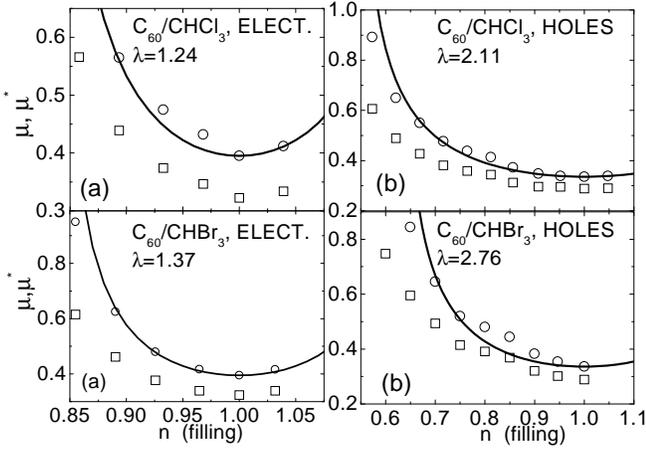}
 \vspace{-2mm}\caption{\small{Coulomb pseudopotential $\mu$ (open
  circle), $\mu^{*}$ (open square) and theoretical fit (solid line) versus filling $n$: (a)
  electron-doped case, (b) hole-doped case.}}\vspace{-1mm}
 \end{figure}
 \begin{figure}[!]
 \vspace{-5mm}
 \includegraphics[keepaspectratio,width=\columnwidth]{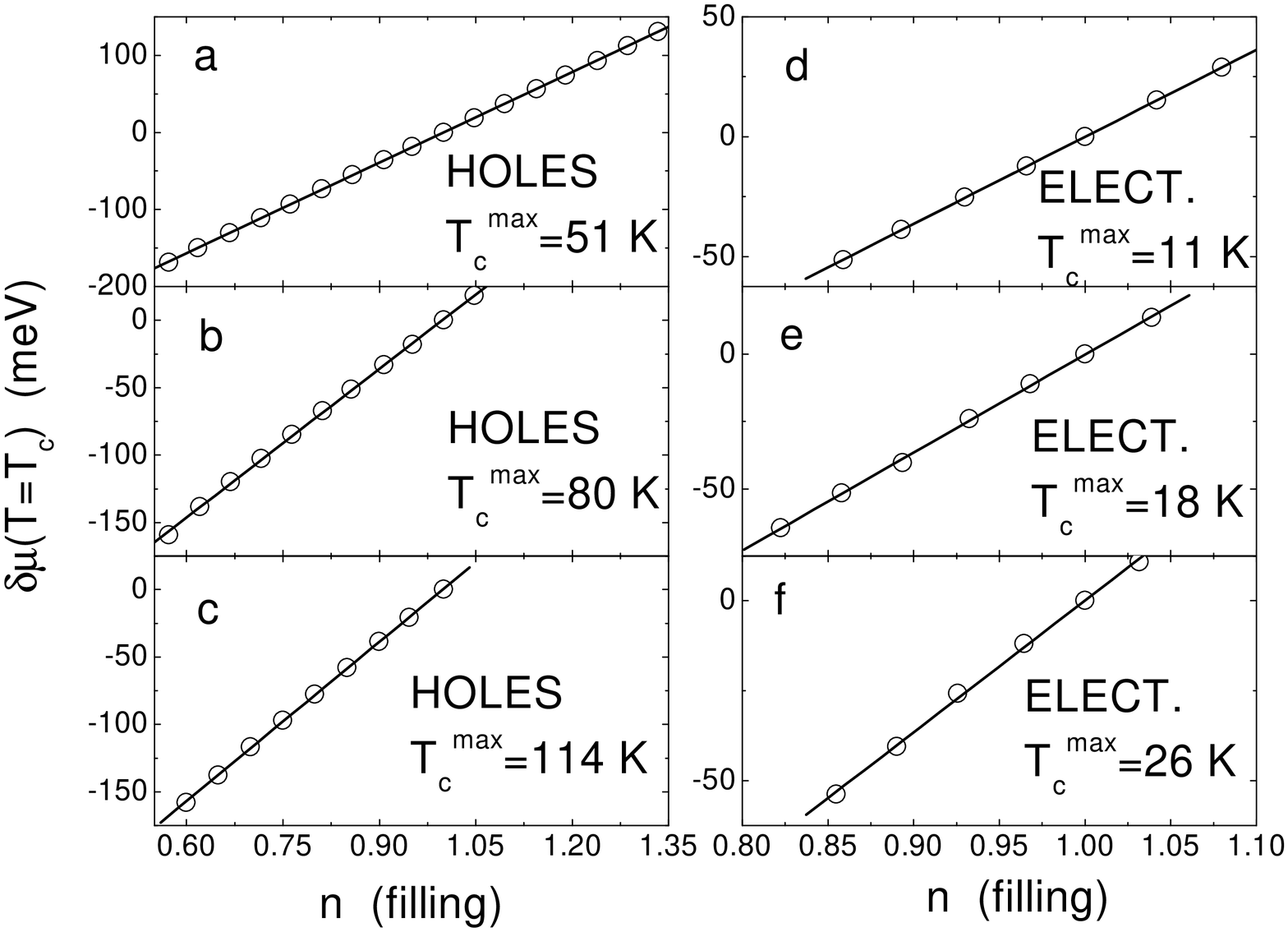}
 \vspace{-5mm}\caption{\small{The chemical potential shift (open circles), at
$T=T_{c}$, calculated from Eliashberg equations and the linear fit
(solid line), in the hole-doped cases (a,b,c) and electron-doped
cases (d,e,f).}}\vspace{-1mm}
 \end{figure}
 \begin{figure}[t]
 \vspace{-5mm}
 \includegraphics[keepaspectratio,width=\columnwidth]{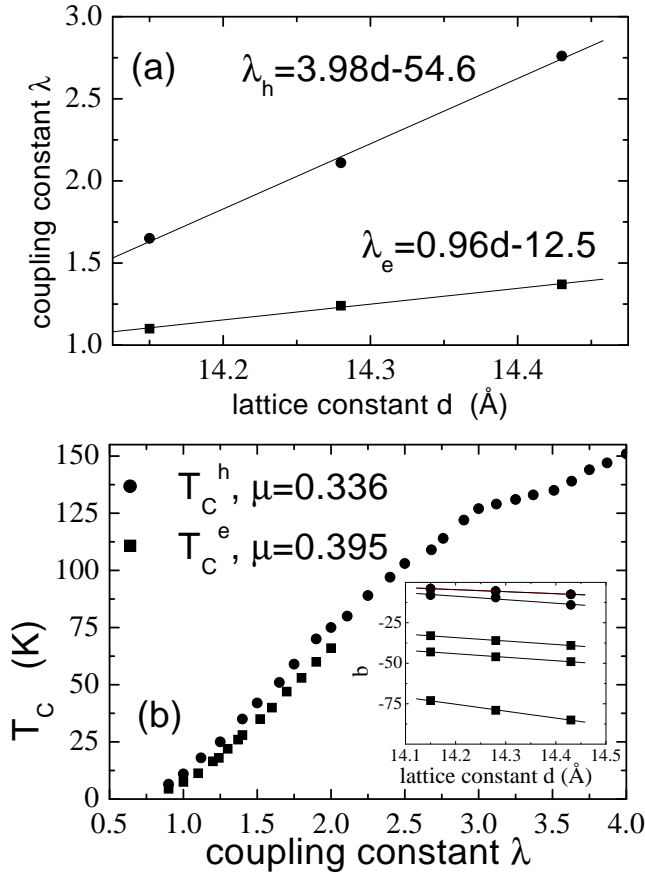}
 \vspace{-10mm}\caption{\small{(a) (a) Coupling constant $\lambda$ versus lattice constant $d$ in the electron-doped case (full squares), hole-doped case (full circles) and their linear fits (solid lines);
  (b) T$_{c}$ versus $\lambda$ in the electron-doped case (full squares) and hole-doped case (full circles);
  in the inset, the parameter $b$ is plotted versus $d$
  in the electron-doped case (full squares) and hole-doped case
  (full circles) together with the linear fits (solid lines).}}\vspace{-1mm}
 \end{figure}
 \begin{table}[!]
\begin{center}
\begin{tabular}{|c|c|c|c|c|}
\hline
  & $C_{60}$ & $C_{60}/CHCl_{3}$ & $C_{60}/CHBr_{3}$ & $C_{60}/?$\\ \hline
  $T_{c,h}^{max} \mathrm{(K)}$ & 51 & 80 & 114 & 135\\ \hline
  $T_{c,e}^{max} \mathrm{(K)}$& 11 & 18 & 26 & 35  \\ \hline
 $\lambda_{h}$ & 1.65 & 2.11 & 2.76& 3.51   \\ \hline
  $\lambda_{e}$ & 1.1 & 1.24 & 1.37 & 1.52 \\ \hline
  $\Delta_{h}\mathrm{(meV)}$ & 9.9 & 16.2 & 25.7 & 35.5 \\ \hline
  $\Delta_{e}\mathrm{(meV)}$ & 1.9 & 3.2 & 4.9 & 6.5  \\ \hline
  $b^{3D}_{h} (a=1.1)$ & -4 & -5.5 & -7.5& -9.6\\ \hline
   $b^{2D}_{h} (a=1.9)$ & -4 & -5.5 & -7.5 & -9.6 \\ \hline
    $b^{2D}_{h} (a=13.8)$ & -8 & -9.5 & -14 & -17 \\ \hline
    $b^{3D}_{e} (a=0.77)$ & -33 & -36 & -39 & -43 \\ \hline
  $b^{2D}_{e} (a=2.1)$ & -43 & -46 & -49 & -53 \\ \hline
  $b^{2D}_{e} (a=10.3)$ & -73 & -79 & -85 & -92 \\ \hline
  $d $(\AA) & 14.15 & 14.28 & 14.43& 14.6 \\ \hline
\end{tabular}
\caption{\small{Important quantities for the six cases examined.
}}\vspace{-2mm}
\end{center}
\end{table}
 In figure 5 the $\mu$ and $\mu^{*}$ values calculated by the solution of Eliashberg equations
 (1)-(4) applied to the $T_{c}(n)$ data of lattice-expanded $C_{60}$ are shown, as well as
 the best fit obtained by using eq. 14 and eq. 17. Also now the fitting curves
 obtained by using the 2D and 3D formula are perfectly superimposed. It is clear that the fits
 are equally good as those shown in Fig. 2.

Unexpectedly, we find that in all cases the shift of the chemical
potential at $T=T_{c}$, calculated by solving the Eliashberg
equations, is a linear function of the filling as we can see in
the six panels of figure 6. The reason of this behaviour will be
the subject of further investigation.

 Figure 7 (a) shows that the calculated coupling constant $\lambda$ is a linear function of the
lattice constant d. We can now extrapolate this behaviour to
calculate the coupling constant of a hypothetical material with
lattice constant $d\approx14.6 \AA$ (the increase is of $1\%$ with
respect to the case where $T_{c}=$ 117 K) \cite{refe13}:
$\lambda_{h}=3.51$ and $\lambda_{e}=1.52$. With these $\lambda$
values and, by using the Eliashberg equations, we determine the
corresponding critical temperatures that are 135 K and 35 K,
respectively.
 Finally, figure 7 (b) reports the critical
temperature as a function of the coupling constant, for constant
$\mu$ and $n=1$, in the hole- and electron-doped case. The
critical temperature, in the hole-doped case, does not increase
linearly for $\lambda\gtrsim 3 $ $(d\gtrsim 14.5 \AA)$, in
disagreement with the theoretical predictions of Ref. 12. On the
contrary, $T_{c}$ drops to zero for $\lambda\lesssim 0.8$
$(d\lesssim 13.9 \AA)$ as reported in Ref. 20.

In the inset of figure 7 (b) we can observe that also the
parameter b is a linear function of the lattice constant d.
 All the results are finally summarized in table I. In principle, the predictions of Fig. 7(b)
and of the last column of Table I concerning $T_{c}$ and the gap
could be experimentally verified in the future. As far as the
lattice constant dependence of $T_{c}$ shown in Table I is
concerned, we admit to that the $T_{c}$ enhancement might be not
due to the increase of the lattice constant but due to the
enhanced polarization and the resulting enhanced screening of the
Coulomb interaction. Unfortunately, the formalism proposed here is
not powerful enough to discriminate between both scenarios.

  In conclusion, the experimental results concerning the complete doping
  dependence of $T_{c}$ in field-effect hole-doped and electron-doped C$_{60}$
  can be naturally explained by the Eliashberg theory
   generalized to the case where the conduction band is finite, non half-filled and the
Coulomb pseudopotential is filling-dependent in a very simple and
physically reasonable way. All the values of the physical
parameters of the present model appear plausible.

Many thanks are due to O.V. Dolgov, E. Cappelluti, C. Grimaldi and
D. Daghero for the useful discussions.

 \end{document}